\documentclass[aps,pra,twocolumn,superscriptaddress,floatfix]{revtex4-1}
\usepackage{color}
\usepackage{amsmath,amsfonts,amssymb}
\usepackage{amsthm}
\usepackage{bm}
\usepackage{graphicx}
\usepackage{subfigure}
\usepackage{bbm}
\usepackage{epstopdf}
\usepackage{epsfig}
\usepackage{verbatim}
\usepackage{array}
\usepackage{ulem}
\usepackage{notes2bib}
\usepackage[T1]{fontenc}
\usepackage{tgtermes}

\usepackage{dcolumn}
\usepackage{braket}
\usepackage[unicode=true,bookmarks=true,bookmarksnumbered=false,bookmarksopen=false, breaklinks=true,pdfborder={0 0 1},backref=false,colorlinks=true]{hyperref}
\hypersetup{linkcolor=magenta, urlcolor=blue, citecolor=blue, pdfstartview={FitH}, unicode=true}
\begin{document}

\title{Zeeman field induced corner states in the Kane-Mele-Hubbard model}

\author{Jie Zhang}
\affiliation{School of Physics and Technology, Wuhan University, Wuhan 430072, China}
\author{Han Xu}
\affiliation{School of Physics and Technology, Wuhan University, Wuhan 430072, China}
\affiliation{Computational Materials Science Research Team, RIKEN Center for Computational Science (R-CCS), Kobe, Hyogo, 650-0047, Japan}
\author{Yu Wang}
\email{yu.wang@whu.edu.cn}
\affiliation{School of Physics and Technology, Wuhan University, Wuhan 430072, China}

\date{\today}

\begin{abstract}
We investigate how the in-plane Zeeman field and the Hubbard interaction can jointly affect the topological states in the Kane-Mele-Hubbard model. At low Zeeman field, the projector quantum Monte Carlo (PQMC) simulations demonstrate Mott transitions between the higher-order topological insulator induced by the in-plane Zeeman field and the antiferromagnetic insulator induced by the Hubbard interaction. The interacting higher-order topological state on a finite-sized honeycomb lattice is characterized by the local density of states derived from the real-space spectral function. In the higher-order topological phase, the mirror-inversion symmetry protected corner states exist on the diamond-shaped honeycomb lattice. By comparing the spatial distribution of corner states between the Kane-Mele and Kane-Mele-Hubbard models, we show that the effect of the Hubbard interaction is to contribute some extra in-plane Zeeman field. At the mean-field level, we calculate the full phase diagram of the Kane-Mele-Hubbard model under the influence of the in-plane Zeeman field. The mean-field phase diagram shows that a high in-plane Zeeman field drives the system into a spin-polarized (topologically trivial) phase. In particular, the upper limit of the Zeeman field that can induce corner states in the Kane-Mele model is found to be $h_c= 1.0(3)$. In the parameter region of PQMC simulations, the mean-field solutions are in fair agreement with the PQMC results.
\end{abstract}

\maketitle

\section{\label{sec:introduction}Introduction}
In recent years, a new class of topological insulators (TIs) termed as higher-order topological insulators (HOTIs), have emerged as an active topic in topological physics \cite{wladimir2017quantized,benalcazar2017electric,song2017d,van2018higher,benalcazar2019quantization,xie2021higher}. Unlike conventional (first-order) TIs owning gapless modes at $(d - 1)$-dimensional boundaries in $d$ dimensions \cite{kane2005z2,fu2007topological,fu2007inversion,hasan2010colloquium,qi2011topological}, the $n$th-order TIs feature symmetry-protected gapless modes at $(d - n)$-dimensional boundaries ($1<n\leqslant d$) \cite{frank2018higher}. Much research has been driven by efforts to seek HOTIs in two- and three-dimensional topological materials.

It has been proposed that three-dimensional (3D) first-order TIs can be turned into HOTIs by imposing a magnetic field or inducing magnetization on their surfaces, since the magnetic stimulation can open a gap in Dirac surface states, leaving 1D gapless hinge states unaffected \cite{sitte2012topological,zhang2013surface}. From a different perspective,  quantized multipole (e.g. quadrupole and octupole) insulators that host gapless states at corners or hinges have also been predicted theoretically  \cite{wladimir2017quantized,benalcazar2017electric,guo2022quadrupole}. The search for HOTIs has now been extended into various systems, including semimetals \cite{ezawa2018highera,lin2018topological,ghorashi2020higher,wang2020higher}, superconductors \cite{song2017d,khalaf2018higher,hsu2018majorana}, graphene \cite{sheng2019two,lee2020two}, quasi-crystal systems \cite{varjas2019topological,agarwala2020higher,chen2020higher,spurrier2020kane,hua2020higher}, and non-Hermitian systems \cite{kunst2018biorthogonal,liu2019second,ghatak2019new}. More excitingly, HOTIs have been experimentally realized in 3D TIs such as bismuth \cite{schindler2018higher} and bismuth bromide \cite{noguchi2021evidence}, as well as in 3D superconductors such as $\textrm{WTe}_2$ \cite{choi2020evidence}. Although HOTIs are yet to be found experimentally in 2D materials, it has been proposed theoretically that imposing an in-plane Zeeman field on 2D first-order TIs may induce corner states, converting first-order TIs into HOTIs  \cite{ezawa2018topological,ren2020engineering,chen2020universal}. This idea is promising based on the extensive inventory of 2D first-order TIs.

In the meantime, the effects of electron-electron interactions in HOTIs have also been studied theoretically, trying to answer how higher-order topological states are influenced by interactions. Exact diagonalization calculations show that HOTI phases in 3D second-order TIs \cite{li2022green} and the quantized multipole insulators \cite{araki2020zq} are stable against weak interactions. At strong coupling, quantum Monte Carlo (QMC) simulations demonstrate a Mott transition from the HOTI to the topologically trivial antiferromagnetic (AFM) insulator \cite{peng2020correlation}. In these studies, the characterization scheme for the higher-order topological phases of interacting HOTIs is to evaluate the topological invariants such as magneto-electric polarizations \cite{li2022green}, $\mathbb{Z}_Q$ Berry phase \cite{araki2020zq} and Wannier-band polarizations \cite{peng2020correlation}) in terms of the zero-frequency single-particle Green's functions. However, the topological invariant calculated with zero-frequency Green's functions remains the same for both the topological phase and the topologically trivial Mott insulator, and thus fails to describe the topological Mott transition \cite{he2016topological,zhao2023failure}.

The momentum-space spectral function provides another scheme for characterizing the first-order topological phase of interacting TIs, which has been employed in the QMC simulations of interacting first-order TIs, giving a unified description of the topological Mott transition \cite{hohenadler2011correlation,hohenadler2012luttinger,bercx2014kane}. Obviously the momentum-space spectral function is wholly absent from 2D second-order topological states (corner states) emerging in a finite-sized lattice system with the open boundary condition. How to extend the spectral function approach to the characterization of higher-order topological states is still an open question. In this paper, we propose that the local density of states (DoS) derived from the real-space spectral function can be used to characterize corner states. We shall employ the stochastic analytical continuation method \cite{johan2016a,shao2023progress} to obtain the real-space spectral function from the unequal-time Green's function, thereby deriving the local DoS of the 2D HOTI. We apply this real-space spectral function approach to the QMC study of the Kane-Mele-Hubbard (KMH) model under an in-plane Zeeman field, exploring the interaction effect on corner states.

The rest of this paper is organized as follows. In Sec. \ref{sec:model and method}, the model Hamiltonian, parameters of QMC simulations and the spectral function scheme for the characterization of interacting TI states are introduced. In Sec. \ref{sec:phase diagram}, the phase diagram of the model is studied by QMC simulations and mean-field calculations. The edge states on an armchair ribbon are described by the spectral function in Sec. \ref{sec:CTI}. Subsequently in Sec. \ref{sec:corner_state}, the corner states on a diamond-shaped lattice are characterized by the local DoS derived from the real-space spectral function. In Sec. \ref{sec:int effect}, the interaction effect on HOTIs is investigated. The conclusions are drawn in Sec. \ref{sec:conclusion}.

\section{\label{sec:model and method}Model and method}
\subsection{\label{subsec:GKMH model}Kane-Mele-Hubbard model under an in-plane Zeeman field}
We consider the half-filled KMH model, subject to an in-plane Zeeman field. The total Hamiltonian of the system is given by $H=H_\textrm{KMH}+H_\textrm{Zeeman}$,
\begin{eqnarray}
&&H_\textrm{KMH}=-t\sum_{\langle i,j\rangle}c_{i}^\dagger c_{j}+i\lambda\sum_{\langle\langle i,j\rangle\rangle}c_i^\dagger \nu_{ij}s_{z} c_j\notag\\
&&\qquad\qquad+U\sum_i\left[n_{i\uparrow}-\frac 12\right] \left[n_{i\downarrow}-\frac 12\right],\notag\\
&&H_\textrm{Zeeman}=h\sum_{i}c_i^\dagger s_{x} c_i.
\label{eq:hamiltonian}
\end{eqnarray}
Here $c^\dagger_{i}=(c_{i\uparrow}^\dagger, c_{i\downarrow}^\dagger)^T$ and $s_{\alpha}$ ($\alpha = x, z$) are spin Pauli matrices. The KMH Hamiltonian includes the terms of the nearest-neighbor hopping, the intrinsic spin-orbit coupling (SOC) and the Hubbard interaction. The next-nearest-neighbor hopping amplitude $\lambda$ reflects the strength of SOC, while $\nu_{ij}\bold{\hat {z}}=\boldsymbol{\delta_i}\times \boldsymbol{\delta_j}/|\boldsymbol{\delta_i}\times \boldsymbol{\delta_j}|$ determines the sign. The vectors $\boldsymbol{\delta}_{i} (i=1,2,3)$ connect three nearest neighbors of a lattice site. The in-plane Zeeman term $H_{\textrm{Zeeman}}$ arises from an in-plane magnetic field or magnetic atom doping \cite{cui2013experimental,weng2015quantum}. For convenience, the in-plane Zeeman field is chosen to be $h\bold{\hat{x}}$.

The KMH model is a prototype model for interacting TIs. QMC studies of the KMH model \cite{hohenadler2011correlation,zheng2011particle} predict that increasing Hubbard $U$ can induce a topological Mott transition from the first-order TI into the AFM insulator. In the non-interacting limit $(U=0)$, the KMH model reduces to the Kane-Mele (KM) model. In the presence of an in-plane Zeeman field, the time-reversal symmetry of the KM model is broken, and therefore the spin-helical edge modes along the zigzag boundary vanish. However, the preserved mirror-inversion symmetry $\mathcal{M}_y=\tau_x\otimes (-is_x)$ ($\tau_x$ is the pseudospin Pauli matrix for the sublattice degrees of freedom) gives rise to gapless edge modes along the armchair boundaries \cite{ren2020engineering}, which is the characteristic feature of the 2D crystalline topological insulator (CTI). Meanwhile the mirror-inversion symmetry also ensures the existence of corner states at the intersection of two zigzag edges \cite{ren2020engineering}, which indicates the presence of nontrivial second-order band topology. In the KMH model, however, it remains unknown to what extent the CTI and HOTI phases are affected by the interaction.
\subsection{\label{subsec:spectral function}Spectral functions}
Physics of interacting TIs is beyond the single-particle band picture. For interacting first-order TIs, the spectral function calculated in momentum space can convincingly demonstrates the gapless boundary states, reflecting the existence of the first-order topological phase \cite{hohenadler2011correlation,hohenadler2012luttinger,bercx2014kane}. The momentum-space spectral function is related to the unequal-time Green's function
\begin{equation}
  G(k,\tau)=\int_{-\infty}^{+\infty}\textrm{d}\omega\,\theta(\omega)e^{-\tau\omega}A(k,\omega),
\label{eq:sepcfunc}
\end{equation}
where $\theta(\omega)$ is the step function. Instead of the maximum entropy method \cite{hohenadler2011correlation,hohenadler2012luttinger,bercx2014kane}, we shall employ the stochastic analytical continuation algorithm \cite{johan2016a,shao2023progress} to evaluate the spectral function from unequal-time Green's functions, which is unbiased, independent of tuning parameters, and free from the over-fitting problem.

In 2D HOTIs, gapless corner states emerge on a finite-sized lattice with the open boundaries, in which the momentum-space spectral function is nonexistent. In order to characterize corner states in 2D interacting HOTIs, we propose to calculate the local DoS from the real-space spectral functions, displaying directly the spatial distribution of the corner states. The procedure is described as follows. First, the unequal-time Green's functions $G_{i,j}(\tau)$ are calculated with the projector QMC (PQMC) method. Secondly, utilizing the Mishchenko's stochastic optimization algorithm \cite{johan2016a}, the real-space spectral functions are evaluated in terms of the relation
\begin{equation}
  G_{i,j}(\tau)=\int_{-\infty}^\infty\textrm{d}\omega\,\theta(\omega) e^{-\tau\omega}A_{i,j}(\omega),
\label{eq:local spectral function}
\end{equation}
by performing the unbiased analytical continuation of unequal-time Green's functions.
Then the local DoS at site $i$ can be obtained from the diagonal spectral function
\begin{equation}
  \rho_{i}(\omega)=\frac{1}{2\pi}A_{i,i}(\omega).
  \label{eq:ldos}
\end{equation}
Normalizing the zero-frequency local DoS $\rho_{i}(\omega=0)$ for all lattice sites can give the spatial distribution of corner states (see Section \ref{sec:corner_state}).

\subsection{\label{subsec:pqmc method}Parameters of PQMC simulations}
The PQMC method \cite{blankenbecler1981monte,hirsch1985two,assaad2008computational,wang2014competing} is employed to simulate the KMH model under an in-plane Zeeman field. The in-plane Zeeman field breaks particle-hole symmetry, leading to the sign problem in QMC simulations. As discussed in Appendix, we fine-tune the simulation parameters and find the range of parameters: the in-plane Zeeman field $h \leqslant 0.4$, the coupling strength $U \leqslant U_c$ ($U_c$ is the critical coupling for the Mott transition), the lattice size $L \leqslant 12$, and the projection length $\beta \leqslant 30$, within which the impact of sign problem is mild.

In this work, PQMC simulations are conducted on 24 CPU cores, with 500 warm-up Monte Carlo steps and 500 measurement steps on each core. The Trotter decomposition parameter is set to be $\Delta\tau = 0.1$. The measurements of physical observables are performed near $\beta/2$ during the projection processes. In our PQMC simulations, the nearest-neighbor hopping amplitude $t=1$ is set as the energy unit; the strength of SOC is set to be $\lambda=0.1$.

\section{\label{sec:phase diagram} phase diagram}
In this section, we employ the PQMC method and the mean-field theory to separately calculate the phase diagram of the KMH model under an in-plane Zeeman field. PQMC simulations can accurately determine a small region of the phase diagram corresponding to the control parameters that incur mild sign problem. As complementary, mean-field calculations can construct a full phase diagram of the KMH model.
\subsection{\label{subsec:QMC phase diagram} PQMC simulations}
Since the SOC term reduces the SU(2) symmetry to the U(1) spin-rotational (around the $z$-axis) symmetry, the AFM ordering is confined to the $x$-$y$ plane \cite{rachel2010topological}. Furthermore, the imposed Zeeman field $h\bold{\hat{x}}$ breaks the U(1) symmetry and thus suppresses the formation of antiferromagnetism along the $\bold{\hat{x}}$ direction. Consequently, the long-range $\mathrm{N\acute{e} el}$ order is expected to develop only along the $\bold{\hat{y}}$ direction. For the description of antiferromagnetism, the AFM structure factor along the $y$-axis is defined at the $\Gamma$ point:
\begin{equation}
  S_{{\text{AF}}}^{yy}=\frac{1}{N}\left\langle\left[\sum_{i}\epsilon^{i}S_i^y\right]^2 \right\rangle,
\label{eq:AFSF}
\end{equation}
where $N$ is the total number of sites; $S_i^y$ is the spin component along the $y$-axis at site $i$; $\epsilon^{i}=\pm1$ for site $i$ belonging respectively to the sublattices A and B.

\begin{figure}
  \includegraphics[width=\linewidth]{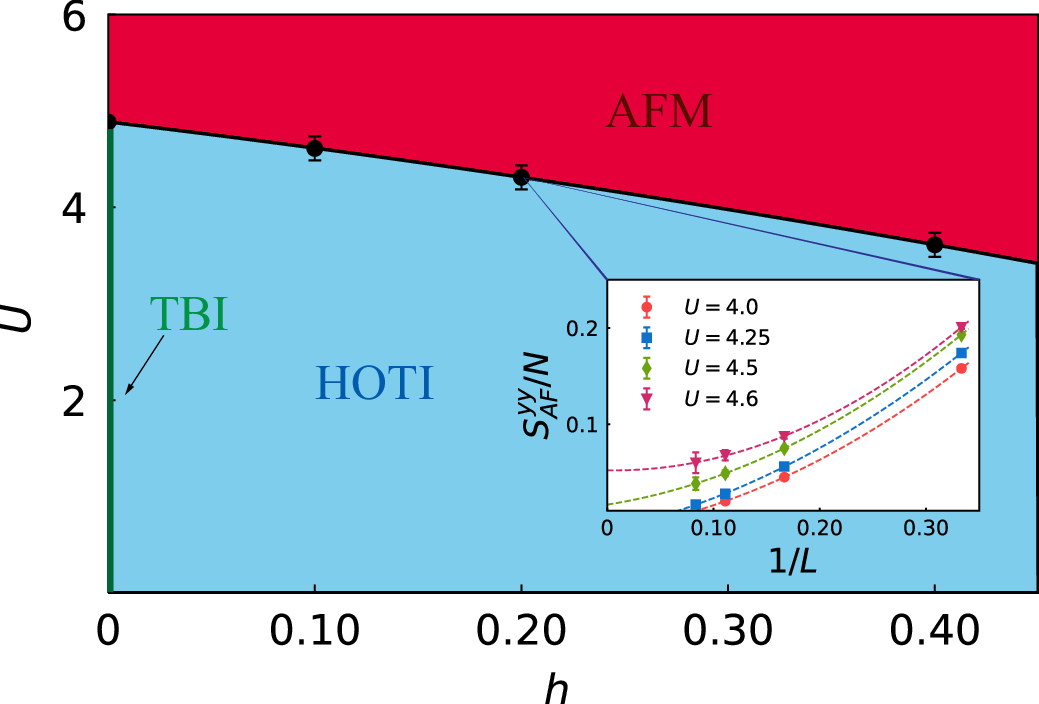}%
  \caption{\label{fig:phasediagram}The PQMC phase diagram of the KMH model under an in-plane Zeeman field. The control parameters lie in the region where the sign problem is mild.     The inset illustrates the finite-size extrapolation of the $y$-axis AFM structure factor for various Hubbard $U$, in the presence of the in-plane Zeeman field $h=0.2$.}
\end{figure}

We employ PQMC simulations to calculate the $y$-axis AFM structure factor for various Hubbard interactions $U$ and Zeeman fields $h$ on the honeycomb lattices with linear sizes $L=3,6,9,12$ under the periodic boundary condition. For given values of $U$ and $h$, the $L\rightarrow\infty$ limit of the AFM structure factor can be obtained by finite-size scaling. The inset plots of Fig.~\ref{fig:phasediagram} display curves for finite-size extrapolation of the $y$-axis AFM structure factor (a quadratic polynomial fitting $a+b/L+c/L^2$ is used) when the Hubbard interaction $U=4, 4.25, 4.5, 4.6$, and the magnitude of the Zeeman field $h=0.2$. It is seen that the onset of AFM ordering occurs in the small coupling interval $4.25 <U_c <4.5$ at $h=0.2$, and accordingly the critical coupling $U_c$ is estimated to be the midpoint value of this small interval.

Figure~\ref{fig:phasediagram} presents the phase diagram of the KMH model under an in-plane Zeeman field. The $h$-dependence of $U_c$ determines the phase boundary, separating the AFM and non-AFM phases. In the absence of the Zeeman field, the Mott transition from the topological band insulator (TBI) phase to the AFM phase occurs at $U_c \approx 4.9$ in the KMH model \cite{hohenadler2011correlation,zheng2011particle}. In the presence of the Zeeman field, the non-AFM phase is found to be a HOTI as discussed in Sec. \ref{sec:corner_state}.

\subsection{\label{subsec:mean-field phase diagram} Mean-field calculations}
Due to the sign problem, PQMC simulations can only determine the partial region of the phase diagram. We now analyze the unknown region of the phase diagram that is beyond the reach of QMC simulations. In the large-$U$ limit, the energy scale of the AFM order is $J \sim 1/U$. Thus the spin polarization caused by the in-plane Zeeman field wins over the AFM ordering at strong coupling, from which one can infer the existence of a phase boundary separating the AFM and disordered phases. In the non-interacting limit (i.e. the KM model), when the Zeeman field $h$ is sufficiently high, the strong spin polarization can disable the intrinsic SOC effect, leading to the topologically trivial band structure. Thus there exists a phase boundary separating the HOTI and spin-polarized disordered phases. In the following, we shall determine the said phase boundaries by the mean-field theory.

In the mean-field approximation, the Hubbard interaction term can be decoupled in terms of the magnetization $\boldsymbol{m}_A=(m_x, m_y,0)$ and $\boldsymbol{m}_B=(m_x, -m_y,0)$ ($A$ and $B$ denote two sublattices). In reciprocal space, the mean-field Hamiltonian of the KMH model under an in-plane Zeeman field is then expressed as
\begin{equation}
  H_{\mathrm{MF}}=\sum_{\boldsymbol{k}}\sum_{l,\alpha}c_{\boldsymbol{k},l,\alpha}^\dagger h_{\boldsymbol{k}} c_{\boldsymbol{k},l,\alpha}+\frac{3N}{U}(\Delta_x^2+\Delta_y^2),
  \label{H_mf}
\end{equation}
with
\begin{eqnarray}
  h_{\boldsymbol{k}}=&&(h+\Delta_x)\tau_0\otimes s_x+\Delta_y\tau_z\otimes s_y \notag\\
  && + \gamma(\boldsymbol{k})\tau_z\otimes s_z-g(\boldsymbol{k})\tau_x\otimes s_0.
  \label{eq:M_mf}
\end{eqnarray}
Here $c_{\boldsymbol{k},l,\alpha}^\dagger=(c_{\boldsymbol{k},l\uparrow}^\dagger,c_{\boldsymbol{k},l\downarrow}^\dagger)$
($l$ is the sublattice index) and $\Delta_{x(y)}=2/3Um_{x(y)}$. The diagonal term $\gamma(\boldsymbol{k})=2\lambda\left[2\mathrm{cos}(3k_y/2)\mathrm{sin}(\sqrt{3}k_x/2)-\mathrm{sin}(\sqrt{3}k_x)\right]$ and the off-diagonal term $g(\boldsymbol{k})=t\sum_{i=1}^{3} e^{-i\boldsymbol{k}\cdot\boldsymbol{\delta}_i}$ come respectively from the SOC and the nearest-neighbor hopping terms. Minimizing the energy with respect to $m_x$ and $m_y$ yields the mean-field equations
\begin{eqnarray}
  \Delta_x&&=\frac{U}{6N}\sum_{\boldsymbol{k}}\left(\frac{h+\Delta_x+\left|g(\boldsymbol{k})\right|}{q_+(\boldsymbol{k})}+\frac{h+\Delta_x-\left|g(\boldsymbol{k})\right|}{q_-(\boldsymbol{k})}\right),\notag\\
  1&&=\frac{U}{6N}\sum_{\boldsymbol{k}}\left(\frac{1}{q_+(\boldsymbol{k})}+\frac{1}{q_-(\boldsymbol{k})}\right),
  \label{eq:self-cons}
\end{eqnarray}
where $q_\pm(\boldsymbol{k})$ are two eigenvalues of $h_{\boldsymbol{k}}$:
\begin{eqnarray}
  &&q_{\pm }(\boldsymbol{k}) =  \\
  &&-\sqrt{\gamma(\boldsymbol{k}) ^2+(h+\Delta_x)^2+\Delta_y^2+\left|g(\boldsymbol{k})\right|^2\pm 2\left|g(\boldsymbol{k})\right|\left|h+\Delta_x\right|}.\notag
  \label{eq:eig}
\end{eqnarray}

\begin{figure}
  \includegraphics[width=\linewidth]{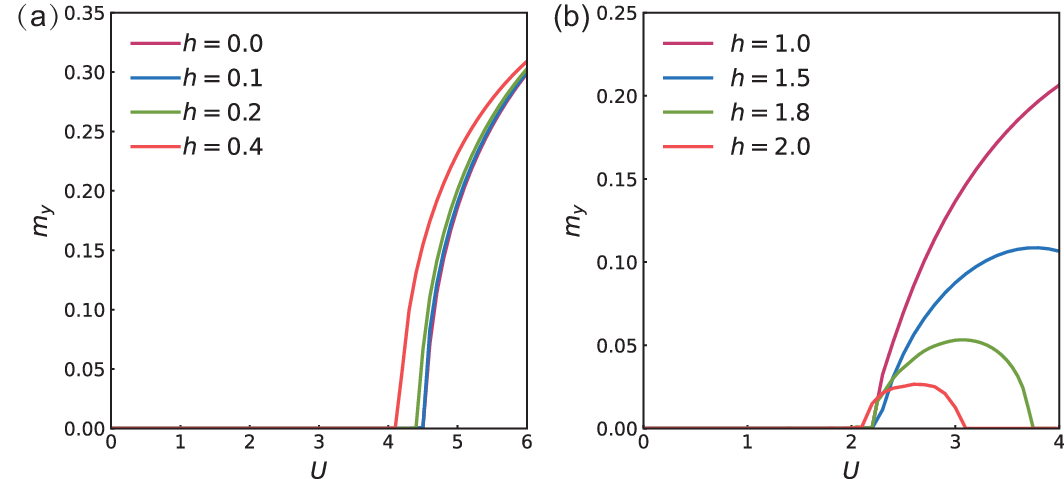}%
  \caption{\label{fig:meanfield_order}The mean-field solution of $m_y$ as a function of $U$ at low (a) and high (b) in-plane Zeeman fields.}
\end{figure}

The AFM order parameter $m_y$ can be calculated self-consistently from Eq.~(\ref{eq:self-cons}). The mean-field solutions of $m_y$ for low and high Zeeman fields are shown respectively in Figs.~\ref{fig:meanfield_order}(a) and \ref{fig:meanfield_order}(b). At high Zeeman field, the AFM order ($m_y\neq0$) is evidently bounded by the lower and upper critical couplings. Therefore, the AFM region is enclosed by the phase boundary that is determined by the self-consistent mean-field calculations of $m_y$.

\begin{figure}
  \includegraphics[width=\linewidth]{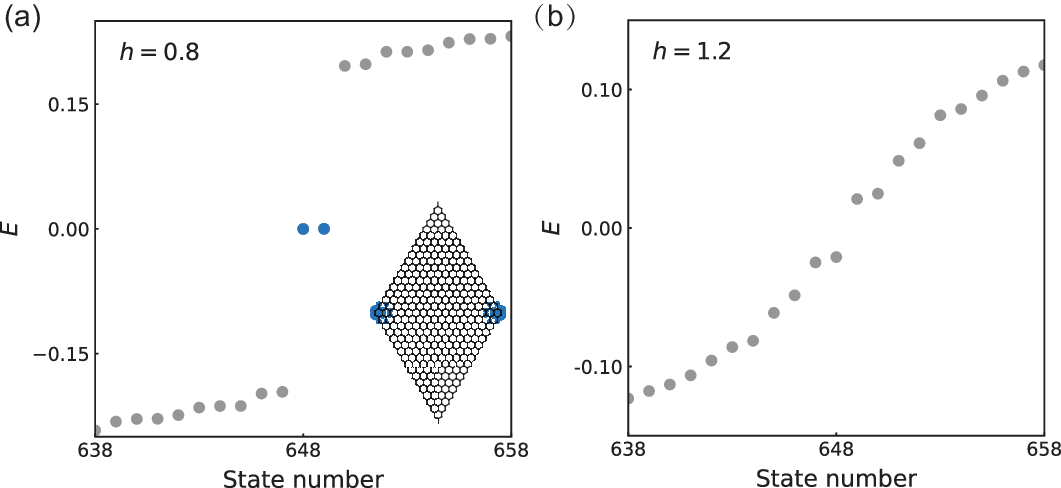}%
  \caption{\label{fig:meanfield_energylv}Energy levels of the mean-field Hamiltonian for (a) $h=0.8$ and (b) $h=1.2$. The inset of (a) illustrates the spatial distribution of the corner states (blue colored). Here the Hubbard interaction $U=1.0$.}
\end{figure}

The topological nature of the mean-field Hamiltonian of the KMH model can be identified via solving its eigen energy levels on a diamond-shaped honeycomb lattice \cite{ren2020engineering,chen2020universal}. If a zero-energy mode is found (in between the energy gap), the system is identified as a HOTI; otherwise, the system is topologically trivial. At $U=1.0$, the eigen energy levels of the mean-field Hamiltonian on a diamond-shaped honeycomb lattice are solved for $h=0.8$ and $h=1.2$, as shown in Figs.~\ref{fig:meanfield_energylv}(a) and \ref{fig:meanfield_energylv}(b), respectively. This indicates that there exists a critical in-plane Zeeman field $h$ (in between 0.8 and 1.2) for inducing the corner states at $U=1.0$. For $U<3$, we calculate the eigen energy levels of the mean-field Hamiltonian for widely varying Zeeman fields $h$, searching for the critical Zeeman field at which the zero-energy mode vanishes. In this way, the phase boundary between the HOTI and disordered (spin-polarized) phases can be determined.

\begin{figure}
  \includegraphics[width=\linewidth]{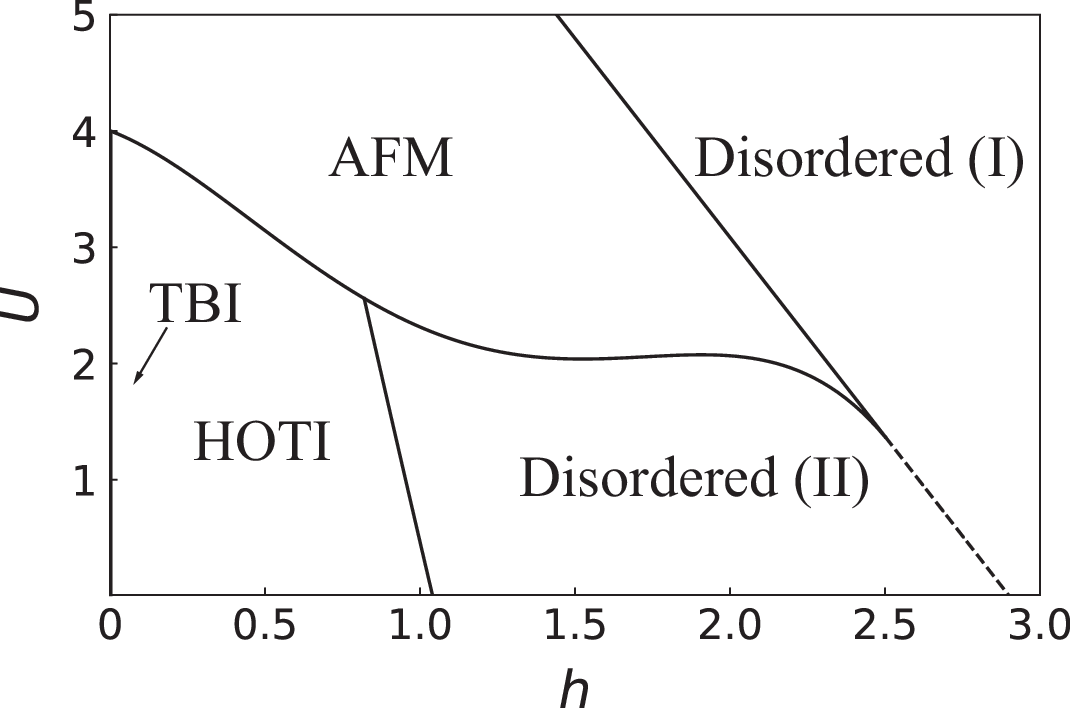}%
  \caption{\label{fig:phasediagram_mf}The mean-field phase diagram of the KMH model under an in-plane Zeeman field. The region of the disordered phase can be divided into two subregions corresponding to the fully spin-polarized states (Disordered (I)) and the partially spin-polarized states (Disordered (II)). The dotted line corresponds to the saturated magnetization $m_{x}=0.5$.   }
\end{figure}

Figure~\ref{fig:phasediagram_mf} shows the mean-field phase diagram of the KMH model under an in-plane Zeeman field. Here the Hubbard interaction $U$ and the applied in-plane Zeeman field $h$ serve as control parameters. Moderate and weak in-plane Zeeman fields can induce the HOTI phase in the KMH model at weak and moderate coupling. In a high Zeeman field, the suppression of the AFM order reproduces the effect that the external-field induced electronic spin polarization can destroy the AFM ordering \cite{bercx2009magnetic,meng2022magnetic}. In the noninteracting limit ($U=0$), we determine the critical in-plane Zeeman field $h_c=1.0(3)$ (at which the HOTI phase vanishes), in agreement with the previous estimation for the KM model \cite{chen2020universal}.

\section{\label{sec:CTI}Edge states on armchair ribbons}
\begin{figure}
  \centering
  \includegraphics[width=\linewidth]{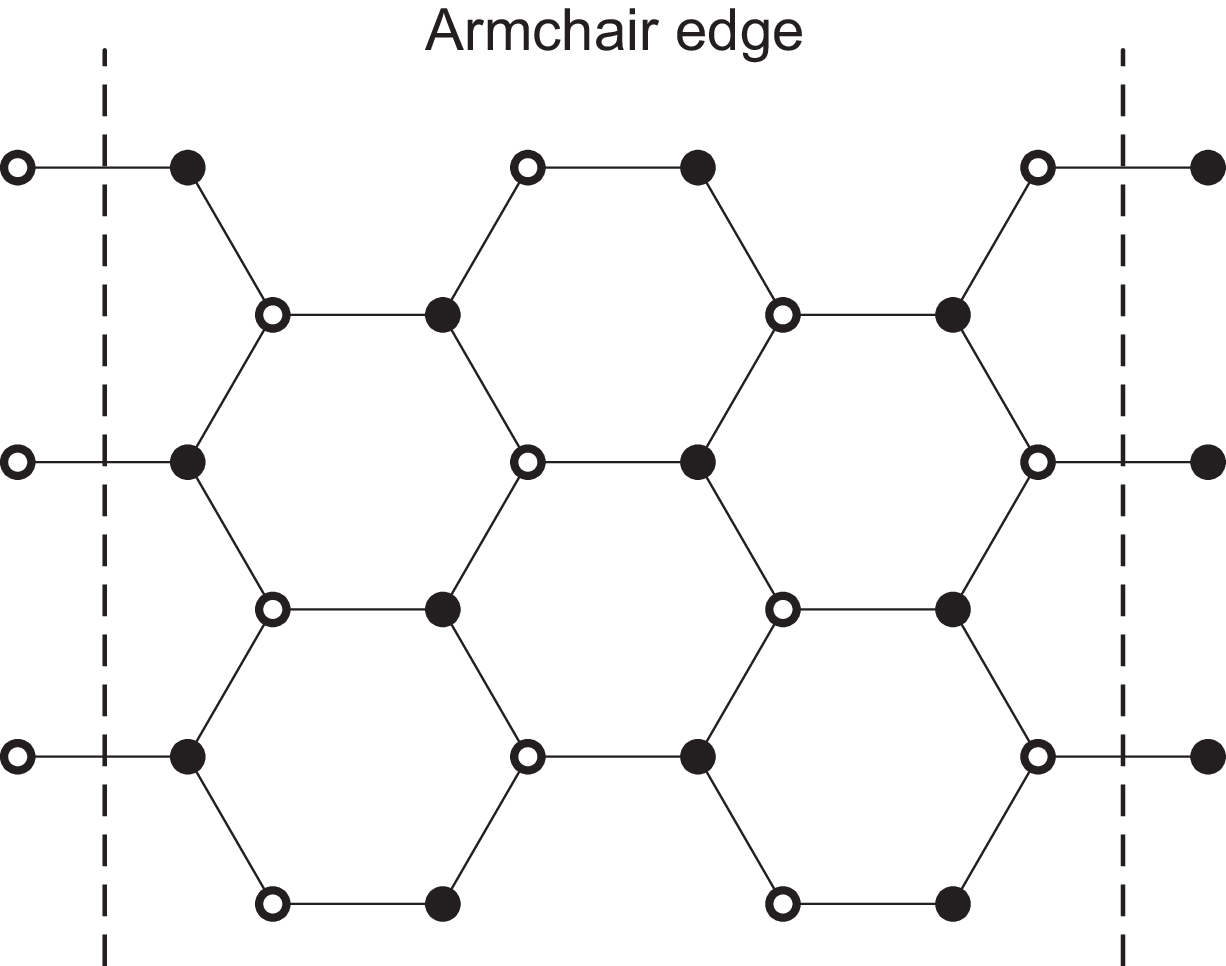}
  \caption{\label{fig:armchair}The honeycomb ribbon with armchair edges. The periodic boundary condition is applied along the $x$-direction, while the open boundary condition is imposed on the $y$-direction.}
  \end{figure}

In this section, we study the influence of the Hubbard interaction on the edge states in the KMH model on the armchair honeycomb ribbon in the presence of an in-plane Zeeman field $h=0.2$. We consider a quasi-1D armchair honeycomb ribbon as depicted in Fig. \ref{fig:armchair}, and calculate the momentum-space spectral function $A(k,\omega)$ by imposing the periodic boundary condition along the $x$-axis and the open boundary condition along the $y$-axis. In our simulation, the size of armchair honeycomb ribbon is $L_{x}\times L_{y}=16\times 8$.
The unequal-time Green's function for the quasi-1D armchair honeycomb ribbon is defined as
\begin{equation}
  G_{\textrm{edge}}(k,\tau)=\frac{1}{L_x} \sum_{i,j\in\textrm{edge}} \langle c_{i}(\tau)c_{j}^\dagger(0)\rangle e^{ik(x_i-x_j)}.
\label{eq:lsp_gap}
\end{equation}
Then the momentum-space spectral function $A(k,\omega)$ can be obtained from Eq.~(\ref{eq:sepcfunc}) by employing the stochastic analytical continuation method \cite{johan2016a,shao2023progress}.

\begin{figure}
  \includegraphics[width=\linewidth]{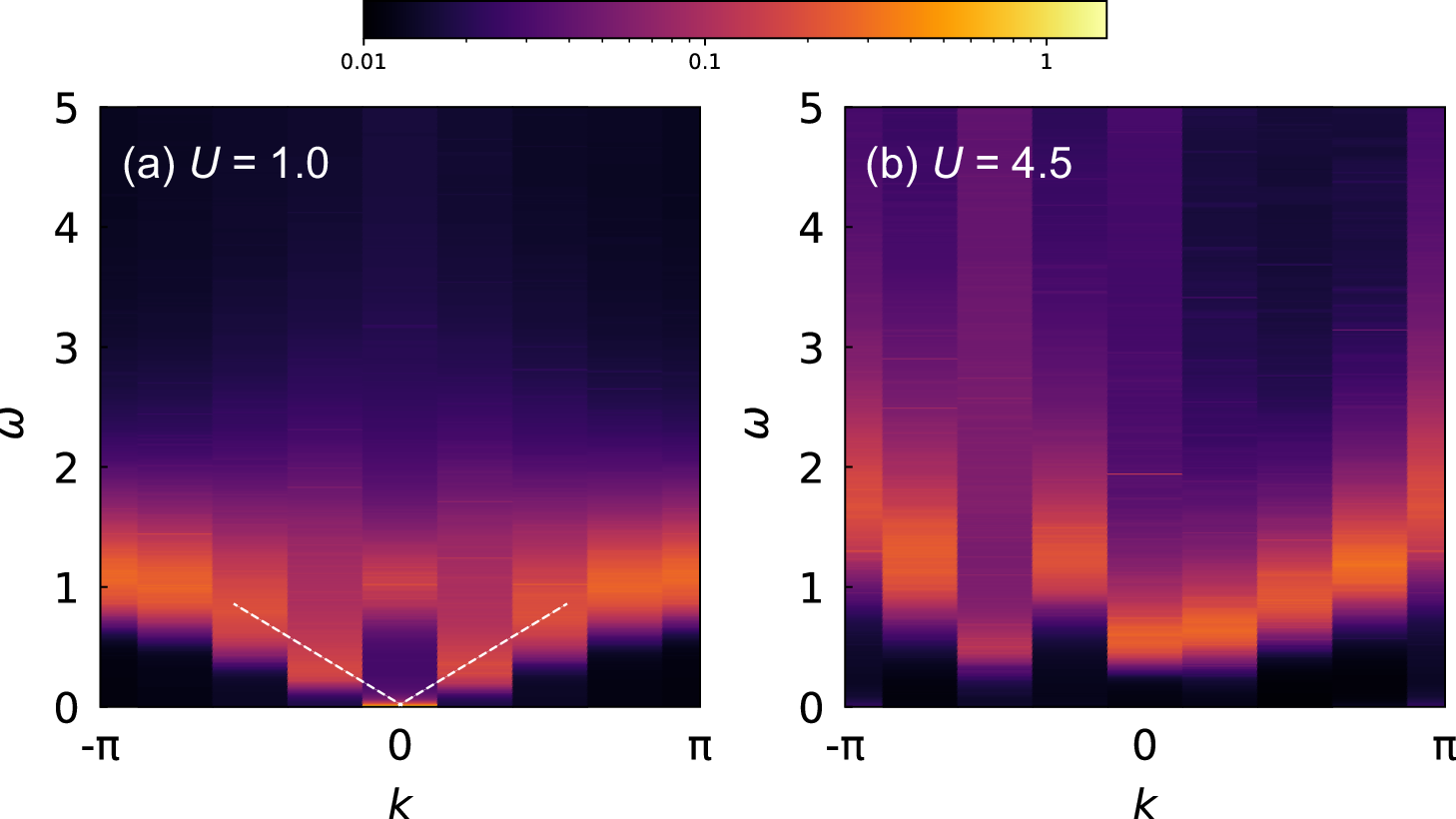}%
  \caption{\label{fig:spec}The momentum-space spectral function $A(k,\omega)$ on the $16\times 8$ armchair honeycomb ribbon describes (a) the gapless edge states at $U=1.0$, and (b) the topologically trivial states at $U=4.5$. Dashed lines in (a) represent the linear dispersion of the edge state spectrum near the $\Gamma$ point.}
\end{figure}

Figure \ref{fig:spec} illustrates the spectral functions $A(k,\omega)$ for two distinct Hubbard interactions. As shown in Fig.~\ref{fig:spec}(a), at $U=1.0$, the gapless edge states emerge at the $\Gamma$ point and the spectral function $A(k,\omega)$ is symmetric about $\Gamma$ point: $A(k,\omega)=A(-k,\omega)$, implying that the system enters the mirror-inversion symmetry protected CTI phase. In contrast, at $U=4.5$, Fig.~\ref{fig:spec}(b) shows that the spectral function $A(k,\omega)$ is asymmetric about $\Gamma$ point and the gapless edge states vanish, indicating that the system enters the Mott-insulating state with spontaneously broken mirror-inversion symmetry. These are in agreement with the PQMC phase diagram obtained in Sec. \ref{subsec:QMC phase diagram}.

\section{\label{sec:corner_state}Corner states on diamond-shaped lattices}
Hosting corner states is a hallmark feature of 2D HOTIs. In this section, we show that the in-plane Zeeman field induces corner states in the KMH model on a diamond-shaped honeycomb lattice. In the non-interacting limit ($U=0$), corner states appear only at the two horizontal corners which are the intersections of two adjacent edges with distinct parity \cite{chen2020universal}. For $U>0$, we calculate the diagonal real-space spectral function $A_{i, i}(\omega)$ and the spatially resolved DoS $\rho_{i}(\omega=0)$ by following the procedure as introduced in Sec. \ref{subsec:spectral function}. In our simulation, the linear size of the diamond-shaped honeycomb lattice is $L=12$.

\begin{figure}
  \centering
  \includegraphics[width=\linewidth]{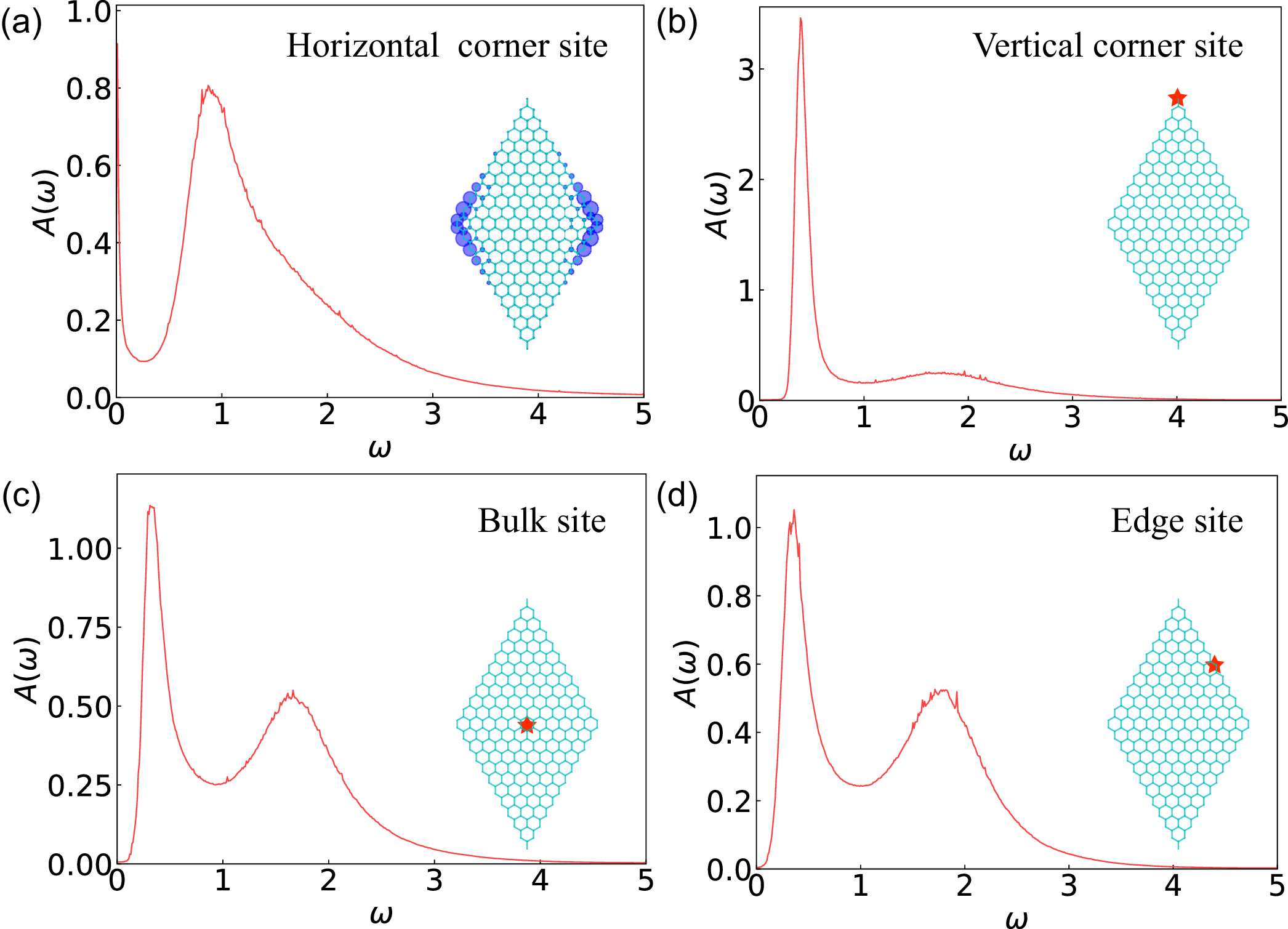}%
  \caption{\label{fig:ldos_l14} The real-space spectral function $A_{i, i}(\omega)$ on the $L=12$ diamond-shaped honeycomb lattice for (a) a horizontal corner site, (b) a vertical corner site, (c) a bulk site, and (d) an edge site. In the inset of (a), the spatial distribution of corner states is depicted. In each inset of (b)-(d), the location of site $i$ is marked by a red star. Here the Hubbard interaction $U=1.0$.}
\end{figure}

\begin{figure}
  \centering
  \includegraphics[width=\linewidth]{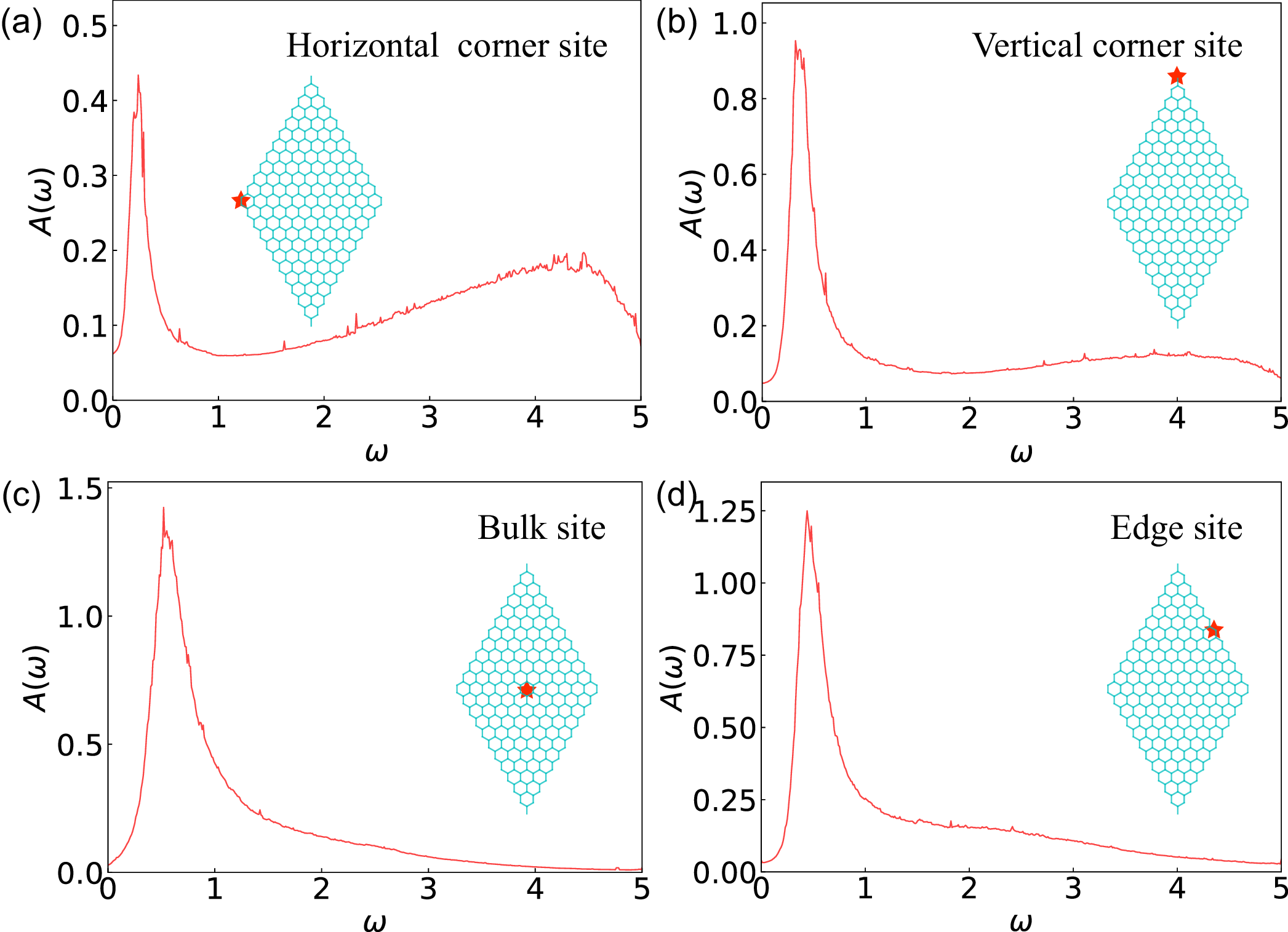}%
  \caption{\label{fig:spec_uc} The real-space spectral function $A_{i, i}(\omega)$ on the $L=12$ diamond-shaped honeycomb lattice for (a) a horizontal corner site, (b) a vertical corner site, (c) a bulk site, and (d) an edge site. The location of site $i$ is marked by a red star in each inset. Here the Hubbard interaction $U=4.5$.}
\end{figure}

At $U=1.0$, Figures~\ref{fig:ldos_l14}(a)-\ref{fig:ldos_l14}(d) show respectively the real-space spectral functions for the sites of the horizontal corner, the vertical corner, the bulk and the edge. At $\omega=0$, the spectral function for a horizontal corner site exhibits a peak, while the spectral functions for other sites reach a trough. Therefore, the spatially resolved DoS $\rho_{i}(\omega=0)$ is only nonzero at around two horizontal corners, characterizing the appearance of corner states. At $U=4.5$, as shown in Figs.~\ref{fig:spec_uc}(a)-\ref{fig:spec_uc}(d), the spectral function for every site reaches a trough at $\omega=0$, indicating that the corner states vanish in the Mott-insulating state. These results demonstrate that the non-AFM phase in the PQMC phase diagram (see Sec. \ref{subsec:QMC phase diagram}) is the HOTI phase.

The zero-frequency real-space spectral function peaking at corner sites intuitively reflects the existence of Zeeman-field-induced corner states in the KMH model, in a format very similar to that in the KM model \cite{ren2020engineering,chen2020universal}. This can be explained based on the understanding of the interaction effect on corner states as discussed in the next section.
\section{\label{sec:int effect}Interaction effect on corner states}
In the thermodynamic limit ($L\rightarrow\infty$), the corner states should be fully localized at the two horizontal corners, corresponding to an extremely sharp peak of the corner-site spectral function at $\omega=0$. However, due to the finite-size effect, the corner states in Fig.~\ref{fig:ldos_l14}(a) are smoothly distributed about a corner, due to the spread of the corner-site spectral function at $\omega=0$. For a finite-size system, the spread of corner states can be used to characterize the effect of the Hubbard interaction and/or the Zeeman field on the corner states \cite{wang2021transport}.

\begin{figure}
  \centering
  \includegraphics[width=\linewidth]{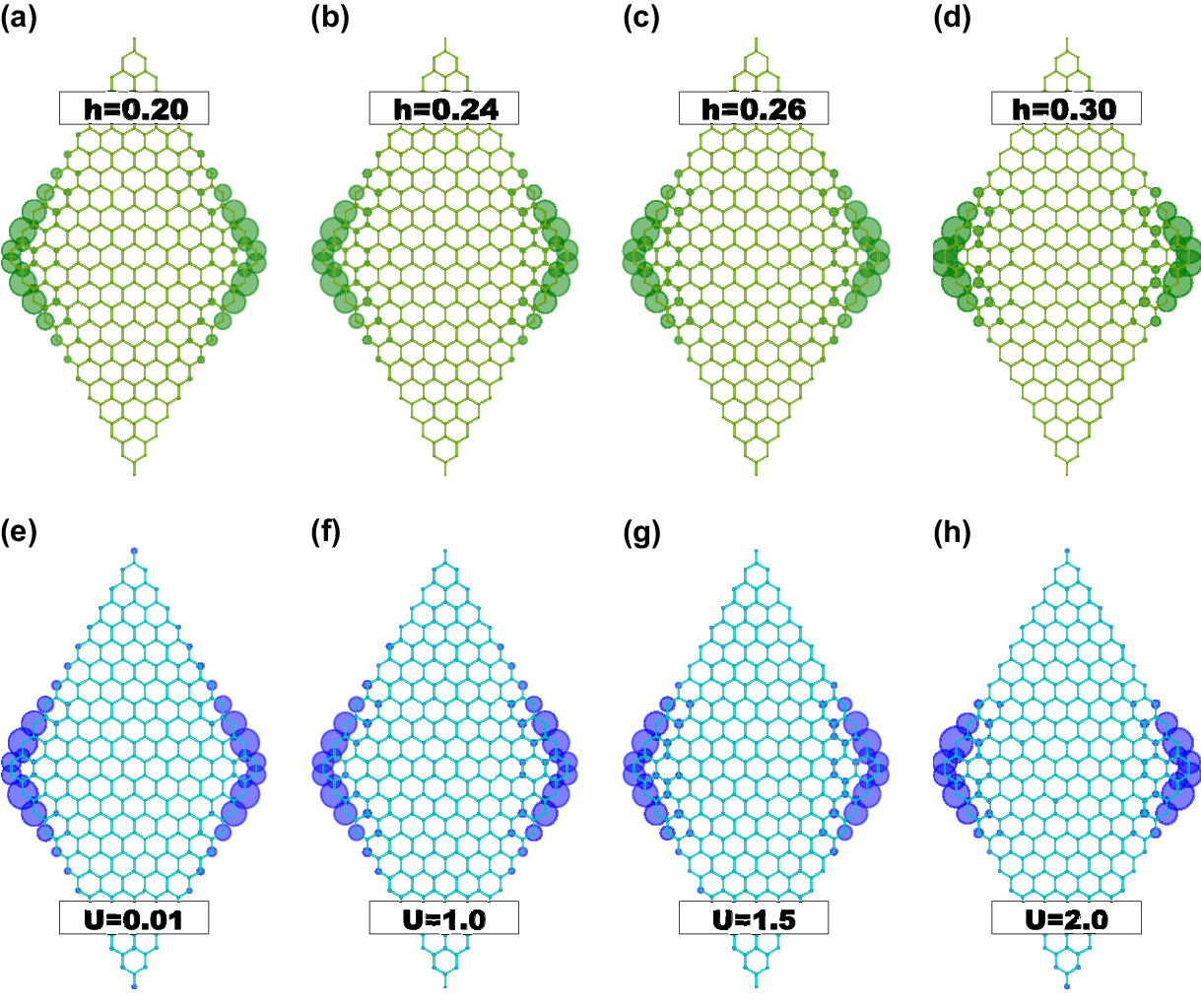}%
  \caption{\label{fig:ldos}(a)-(d) In the KM model, the spread of the corner state distribution narrows with increasing in-plane Zeeman field. (e)-(h) In the KMH model under the in-plane Zeeman field $h_{0}=0.2$, the spread of the corner state distribution narrows with increasing Hubbard interaction. The spatial distributions of corner states in (a)-(d) are identical to those in (e)-(h), respectively.}
\end{figure}

\begin{figure}
  \centering
  \includegraphics[width=\linewidth]{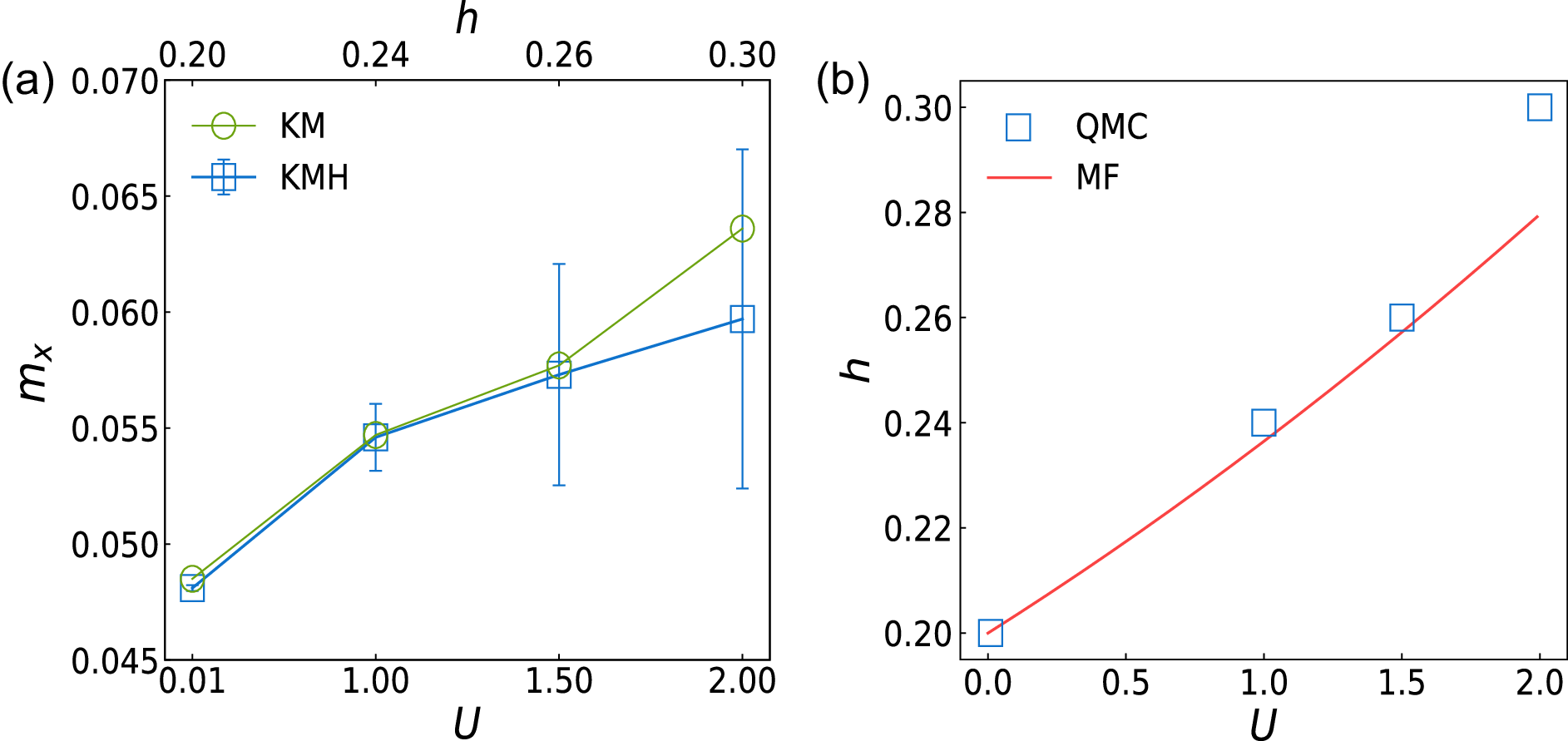}%
  \caption{\label{fig:int_effect}(a) The $x$-axis magnetization $m_x$ in the KM and KMH models induced respectively by the in-plane Zeeman field (top $x$-axis label) and Hubbard interaction (bottom $x$-axis label). (b) Fitting of the correspondence relation $h\leftrightarrow U$ (derived from Fig.~\ref{fig:ldos}) by the mean-field curve $h(U) = h_0 + h_{\mathrm{eff}}(U)$. Here the KMH model is subject to the in-plane Zeeman field $h_{0}=0.2$.}
\end{figure}

Figures~\ref{fig:ldos}(a)-\ref{fig:ldos}(d), acting as a benchmark, show that the spread of the corner state distribution narrows with increasing in-plane Zeeman field in the (noninteracting) KM model. We simulate the distribution of corner states for various Hubbard $U$ in the KMH model under the in-plane Zeeman field $h_{0}=0.2$, and present the results in Figs.~\ref{fig:ldos}(e)-\ref{fig:ldos}(h), showing that the spread of the corner state distribution narrows with increasing Hubbard interaction. Figure~\ref{fig:ldos} indicates that the in-plane Zeeman field can induce the same spread of the corner state distribution in the KM and KMH models, establishing the one-to-one correspondence $h\leftrightarrow U$ between $h$ of the noninteracting case and $U$ of the interacting case. Moreover, the $x$-axis magnetizations $m_x$ in the KM and KMH models are calculated respectively for $h$ and $U$ obeying the correspondence relation $h\leftrightarrow U$ derived from Fig.~\ref{fig:ldos}, yielding almost the same $m_x$ as shown in Fig.~\ref{fig:int_effect}(a). We see that, in the presence of the in-plane Zeeman field, the KMH model behaves like the KM model in the HOTI phase, and the effect of Hubbard $U$ is to contribute some extra Zeeman field to the actual in-plane Zeeman field. Since increasing the in-plane Zeeman field only narrows the spread of the $\omega=0$ corner-site spectral function rather than alters its peak/trough nature, the real-space spectral function of the KMH model thus resembles that of the KM model, as seen in Sec. \ref{sec:corner_state}.

In the half-filled KMH model, charge fluctuations enhance double occupancy of electrons with antiparallel spins on a site, which counteracts magnetization induced by an in-plane Zeeman field. Increasing Hubbard $U$ suppresses charge fluctuations and therefore enhances the Zeeman-field-induced magnetization. In the mean-field picture (see Sec.~\ref{subsec:mean-field phase diagram}), Hubbard $U$ contributes an extra effective Zeeman field $h_{\mathrm{eff}}=2Um_{x}(U)/3$ along the direction of the actual in-plane Zeeman field $h_0$, where the $U$-dependent $x$-axis magnetization $m_{x}(U)$ (mean-field parameter) can be determined self-consistently by the mean-field equations. The mean-field Hamiltonian of the KMH model under a Zeeman field $h_0$ is equivalent to the KM model under a Zeeman field $h(U)=h_{0}+h_{\mathrm{eff}}(U)$. As shown in Fig.~\ref{fig:int_effect}(b), the mean-field $h(U)$ curve is a good fit to the $h\leftrightarrow U$ correspondence relation determined by the PQMC simulations, when $U<2$.


\section{\label{sec:conclusion}conclusions}
We have investigated the higher-order topological phase in the KMH model, subject to an in-plane Zeeman field. In the parameter region where the impact of the QMC sign problem is mild, we perform the PQMC simulations to explore the interaction effects on corner states. We suggest a scheme for a unified description of the higher-order topological states on the finite-sized 2D lattices by using the real-space spectral functions. At weak and moderate coupling, our PQMC simulations demonstrate the existence of edge states on a quasi-1D armchair honeycomb ribbon and of corner states on a diamond-shaped honeycomb lattice. Both the mirror-symmetry protected CTI and HOTI are stable against moderate interactions. Further increasing the Hubbard $U$ leads to a Mott transition to the topologically trivial AFM phase. By comparing the spatial distribution of corner states between the KM and KMH models, we demonstrate that the effect of the Hubbard interaction is to contribute some extra in-plane Zeeman field.

At mean-field level, we establish the complete phase diagram of the Zeeman-field-affected KMH model. At weak and moderate coupling, the Zeeman-field-induced HOTI phase (characterized by the emergence of corner states) exists when the in-plane Zeeman field is weak and moderate. A strong in-plane Zeeman field causes spin polarization, which makes the system topologically trivial. In particular, the upper limit of the Zeeman field for inducing corner states in the KM model is found to be $h_c =1.0(3)$. In the parameter region of PQMC simulations, the mean-field solutions are in fair agreement with the results of PQMC simulations.

In our PQMC study, the corner state is described by the local DoS derived from the real-space spectral function. This scheme can also be implemented in other numerical methods that are capable of computing the real-space unequal-time Green's functions, such as the exact diagonalization and dynamical mean-field theory. The local DoS description of the corner states applies to extensive 2D interacting HOTI models, including interacting quantized multipole insulators \cite{araki2020zq,peng2020correlation}, and offers a useful tool for studying interacting HOTIs.

\acknowledgments
This work is financially supported by the National
Natural Science Foundation of China under Grants No. 11874292, No. 11729402, and No. 11574238.
We acknowledge the support of the Supercomputing Center of Wuhan University.

\appendix
\section*{\label{sec:average sign}Appendix: Average fermion sign}
\begin{figure}
  \includegraphics[width=\linewidth]{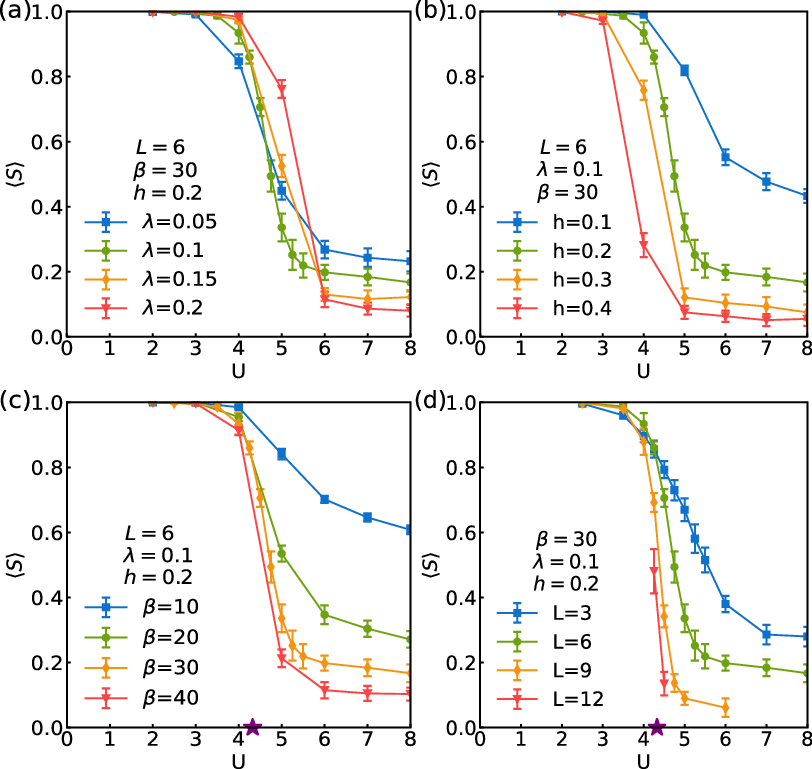}
  \caption{\label{fig:avgsign}Average sign $\braket{S}$ as a function of Hubbard interaction for various (a) spin-orbit couplings $\lambda$, (b) Zeeman fields $h$, (c) projection times $\beta$, and (d) lattice sizes $L$. In (c) and (d), the purple stars on the $x$-axis indicate the critical point $U_c\approx 4.3$ at which the phase transition occurs.}
\end{figure}

In determinant QMC algorithm, the Boltzmann weight $\omega(z)$, which determines the sampling probability of the auxiliary field configuration $z$, might be not positive definite, causing the sign problem. The average sign of the Boltzmann weight $\braket{S}=\sum_z\omega(z)/\sum_z|\omega(z)|$ (ideally $\braket{S}=1$) is a measure of how severe the sign problem becomes \cite{loh1990sign}. We perform QMC simulations of the average sign $\braket{S}$ to determine the parameter region in which the sign problem is mild. Graphs in Fig.~\ref{fig:avgsign} plot the average sign $\braket{S}$ against the Hubbard interaction, under the influence of spin-orbit coupling, Zeeman field, projection time, and lattice size. In the region of simulation parameters, the average sign $\braket{S}$ declines rapidly near critical coupling $U_c$, namely, the sign problem worsens precipitously near quantum critical point \cite{mondaini2022quantum}. Our numerical data reveal the optimal region of simulation parameters: the in-plane Zeeman field $h \leqslant 0.4$, the coupling strength $U \leqslant U_c$, the lattice size $L \leqslant 12$, and the projection length $\beta \leqslant 30$, in which the sign problem is minor.

\vspace{2cm}

% \bibliography{manuscriptNote}
\end{document}